\documentclass[useAMS, usenatbib]{mn2e}
\usepackage[english]{babel}
\usepackage[dvips]{graphicx}
\usepackage{amsmath}
\usepackage{amssymb}
\usepackage{color}
\usepackage{latexsym}
\usepackage{stmaryrd}
\usepackage{times}
\usepackage{url}
\usepackage{hyperref}

\newcommand{\dd}{\mathrm{d}}

\newcommand{\ie}{\textit{i.e.,}~}
\newcommand{\eg}{\textit{e.g.,}~}

\begin{document}

{\setlength\paperheight {297mm}%
\setlength\paperwidth  {210mm}}

\title[Beyond 2nd order convergence in BNS simulations] {Beyond
second-order convergence in simulations of binary neutron stars in full
general-relativity}

\author[D.~Radice et al.]{David Radice$^{1,2}$, Luciano Rezzolla$^{1,3}$ and
Filippo Galeazzi$^{4,1}$ \\
$^1$Max Planck Institute for Gravitational Physics, Albert Einstein
Institute, Potsdam, Germany \\
$^2$Theoretical Astrophysics, California Institute of Technology,
1200 E California Blvd, Pasadena, California 91125, USA \\
$^3$Institut f\"ur Theoretische Physik, Max-von-Laue Str. 1., 60438,
Frankfurt, Germany \\
$^4$Departamento de Astronom\'{\i}a y Astrof\'{\i}sica, Universitat de
Val\`encia, Dr. Moliner 50, 46100, Burjassot (Val\`encia), Spain
}

\maketitle

\begin{abstract}
Despite the recent rapid progress in numerical relativity, a convergence
order less than the second has so far plagued codes solving the
Einstein--Euler system of equations. We report simulations of the
inspiral of binary neutron stars in quasi-circular orbits computed with a
new code employing high-order, high-resolution shock-capturing,
finite-differencing schemes that, for the first time, go beyond the
second-order barrier. In particular, without any tuning or alignment, we
measure a convergence order above three both in the phase and in the
amplitude of the gravitational waves. Because the new code is already
able to calculate waveforms with very small phase errors at modest
resolutions, we are able to obtain accurate estimates of tidal effects in
the inspiral that are essentially free from the large numerical viscosity
typical of lower-order methods, and even for the challenging large
compactness and small-deformability binary considered here. We find a
remarkable agreement between our Richardson-extrapolated waveform and the
one from the tidally corrected post-Newtonian (PN) Taylor-T4 model, with
a de-phasing smaller than $0.4$ radians during the seven orbits of the
inspiral and up to the contact point. Because our results can be used
reliably to assess the validity of the PN or other approximations at
frequencies significantly larger than those considered so far in the
literature, at these compactnesses, they seem to exclude significant
tidal amplifications from next-to-next-to-leading--order terms in the PN
expansion.
\end{abstract}

\begin{keywords}
gravitational waves -- stars: neutron
\end{keywords}

\section{Introduction}

The inspiral and merger of binary neutron stars (BNS) is one of the most
promising sources of gravitational waves (GWs) for future ground-based
laser-interferometer detectors such as LIGO, Virgo or KAGRA
\citep{Sathyaprakash:2009xs}. Because they can travel almost unscattered
through matter, GWs carry valuable information from the deep core of the
neutron stars (NSs) concerning the equation of state (EOS) of matter at
supra-nuclear densities. Unfortunately, they are also extremely hard to
detect, so that their identification and analysis requires the
availability of analytical or semi-analytical GW templates. In turn, the
validation and tuning of these models must be done by matching them with
the predictions of fully non-linear numerical relativity (NR)
calculations, which represent the only means to describe accurately the
late inspiral of BNS \citep{Baiotti:2010, Baiotti2011, Bernuzzi2012,
Hotokezaka2013b}.

While very high-quality NR waveforms of binary black hole mergers are
available, \eg \citep{Aylott:2009ya, Mroue_etal_2013, Hinder2013} (but
see \citep{Zlochower2012}), BNS simulations have been plagued by low
convergence order and large phase uncertainties
\citep{Baiotti:2009gk,Bernuzzi2011}. Furthermore, since NSs have smaller
masses, the merger part of the waveform is out of the frequency band for
the next generation GW detectors, so that EOS-related effects will have
to be most probably extracted from the inspiral signal.  In particular,
EOS-induced effects will be encoded in the de-phasing that the GW signal
will have with respect to the one expected for point particles (PP);
using a post-Newtonian (PN) language, this can be seen as due to the
dissipation of part of the orbital angular energy into tidal deformations
[see, \eg \citet{Damour:2012} for a discussion]. As a result, the measure
of the EOS-induced effects requires very accurate general-relativistic
predictions of the inspiral signal, imposing that this part of the
process is modelled as accurately as possible. Even though accurate
waveforms can be calculated by current codes at very high computational
costs, \citep{Baiotti:2010, Baiotti2011, Bernuzzi2012, Hotokezaka2013b},
their analysis is complicated by the low convergence order of the
employed methods. In particular, the analysis often requires the use of a
time rescaling or alignment of the waves from different resolutions
\citep{Baiotti2011, Hotokezaka2013b}, which is hardly justified from a
mathematical point of view, casting doubts on the robustness of the
results. Finally the goal of exploring accurately a \emph{large} space of
parameters seems out of reach for present fully general-relativistic
codes.

In this \emph{Letter} we show that, by using high-order numerical
methods, it is indeed possible to obtain waveforms for the late-inspiral
of a BNS system of a quality that is almost comparable with the one
obtained for binary black holes~\citep{Hinder2013}, \ie with clean,
higher than second-order convergence in both the phase and the amplitude.

\section{Numerical methods}

The results presented here have been obtained with our new high order,
high-resolution shock-capturing (HRSC), finite-differencing code:
\texttt{WhiskyTHC}, which represents the extension to general relativity
of the \texttt{THC} code \citep{Radice2012a}. In particular, the new code
makes use of the high-order flux-vector splitting finite-differencing
techniques described in \citep{Radice2012a}, but also benefits from the
recent developments in \texttt{Whisky} in terms of the recovery of the
primitive quantities and of the use of tabulated EOSs (see
\citep{Galeazzi2013} for details), although our results will refer to an
ideal-fluid EOS. More specifically \texttt{WhiskyTHC} solves the
equations of general-relativistic hydrodynamics in conservation form
\citep{Banyuls97} using a finite-difference scheme that employs flux
reconstruction in local-characteristic variables using the MP5 scheme
\citep{suresh_1997_amp}. This scheme is formally fifth order in space and
in \citet{Radice2012a} it was shown to lead to a clean fifth-order
convergence in a stringent test involving the propagation of a nonlinear
wave in a flat spacetime.

The spacetime evolution, instead, makes use of the BSSNOK formulation of
the Einstein equations \citep{Nakamura87, Shibata95, Baumgarte99} and it
is performed using the \texttt{Mclachlan} code of the \texttt{Einstein
Toolkit} \citep{Loffler:2011ay, Brown:2008sb, Schnetter-etal-03b} using
a fourth-order accurate finite-difference scheme. To ensure the
non-linear stability of the scheme we add a fifth-order Kreiss-Oliger
type artificial dissipation to the spacetime variables only.

Finally, the coupling between the hydrodynamic and the spacetime solvers
is done using the method of lines and a fourth-order Runge-Kutta
time integrator.

We remark that, with a formal fourth-order of convergence in time and
space, ours is the first higher-than-second order general-relativistic
hydrodynamics code\footnote{Other high-order general-relativistic
hydrodynamic codes have been developed, such as \texttt{WHAM}
\citep{Tchekhovskoy2007} or \texttt{ECHO} \citep{DelZanna2007,
Bucciantini2011}. These codes, however, either use fixed spacetimes or
approximations to general relativity.}.

\section{Binary setup}

The initial data is computed in the conformally flat approximation using
the \texttt{\textsc{Lorene}} pseudo-spectral code \citep{Gourgoulhon01}
and describes two equal-mass NSs in quasi-circular orbit. Its main
properties are summarized in Table \ref{table:models}, and we note that
it is computed using a polytropic EOS with $K = 123.56$ and $\Gamma=2$,
while the evolution is performed using the ideal-fluid EOS with the same
$\Gamma$.

\begin{table}
\caption{\label{table:models} Summary of the considered BNS model. We
  report the total baryonic mass $M_b$, the ADM mass $M$, the initial
  separation $r$, the initial orbital frequency $f_{\mathrm{orb}}$, the
  gravitational mass of each star at infinite separation,
  $M_\infty$, the compactness, $\mathcal{C} = M_\infty/R_\infty$, where
  $R_\infty$ is the areal radius of the star when isolated and the tidal
  Love number, $\kappa_{2}$, \eg \citet{Hinderer09}.}
\vspace{1em}
\resizebox{\columnwidth}{!}{
\begin{tabular}{ccccccc}
  \hline\hline
    $M_b\ [M_\odot]$ &
    $M\ [M_\odot]$ &
    $r\ [\mathrm{km}]$ &
    $f_{\mathrm{orb}}\ [\mathrm{Hz}]$ &
    $M_\infty\ [M_\odot]$ &
    $\mathcal{C}$ & 
    $\kappa_2$ \\
  \hline
  $3.8017$ & $3.45366$ & $60$ & $208.431$ & $1.7428$ & $0.18002$ & $0.05$ \\
  \hline\hline
\end{tabular}
}
\end{table}

Binaries with the same compactness, but different EOS, have also been
considered by \citep{Hotokezaka2013b}, where it was found that
high-compactness binaries are much more challenging to evolve accurately
than low-compactness ones. This is because numerical viscosity becomes
the leading source of de-phasing from the PP limit, since tidal effects
are small. The model chosen here is even more challenging than that in
\citep{Hotokezaka2013b}, as our EOS leads to even smaller tidal
deformabilities (namely, smaller values for the $\kappa_2$ Love number).

All of the runs are performed on a grid covering $0 < x,z \leq 512\
M_\odot$, $-512\ M_\odot \leq y \leq 512\ M_\odot$, where we assume
reflection symmetry across the $(x,y)$ plane and $\pi$ symmetry across
the $(y,z)$ plane. The grid employs six \emph{fixed} refinement levels,
with the finest one covering both stars and we consider three different
resolutions having, in the finest refinement level, a grid spacing of
$h/M_{\odot}=0.25, 0.20$ and $0.14545$, respectively. Our gauges are the
standard $1 + \log$ slicing condition \citep{Bona95b} and the moving
puncture spatial gauge condition \citep{vanMeter:2006vi} with damping
parameter set to $0.3$.

Since our focus is mostly on the accuracy of the methods, we consider the
accuracy of the code by mainly looking at the $\ell = 2, m = 2$ mode of
the Weyl scalar $\Psi_4$ extracted at the fixed coordinate radius of $r =
450\ M_\odot$ ($\simeq 130\ M$). We do not compute the strain as this
involves other uncertainties \citep{Reisswig:2011}, nor we extrapolate in
radius $\Psi_4$ as we expect this not to be a large contribution to our
error budget.  Indeed, for a grid setup similar to ours but for a lower
compactness and smaller total mass model, \citet{Baiotti2011}
estimated a phase uncertainty of $\pm 0.05$ radians, which is negligible
when compared to the uncertainty due to the eccentricity in the initial
data.

\section{Gravitational waves}

The dynamics of the inspiral and merger of BNS has been described many
times and in great detail in the literature, \eg \citep{Baiotti08}; for
this reason we do not give a very in-depth discussion of it here. We only
mention that our stars inspiral for about $8$ orbits before merger and
then rapidly produce a black-hole. For this model, no significant disk is
left behind. The GW signal consists of about $16$ cycles up to merger,
followed by the black hole ringdown.

\begin{figure}
\begin{center}
  \includegraphics[width=\columnwidth]{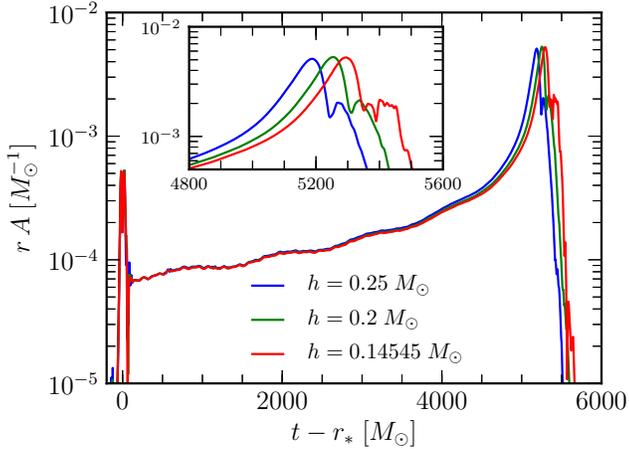}
  \caption{\label{fig:bns60.ampl} Amplitude of the $\ell=2, m=2$ mode of
  the GW Weyl scalar $\Psi_4$ as extracted at radius $r = 450\ M_\odot$
  for three different resolutions.}
\end{center}
\end{figure}

\begin{figure*}
  \begin{minipage}{\columnwidth}
  \begin{center}
    \includegraphics[width=\columnwidth]{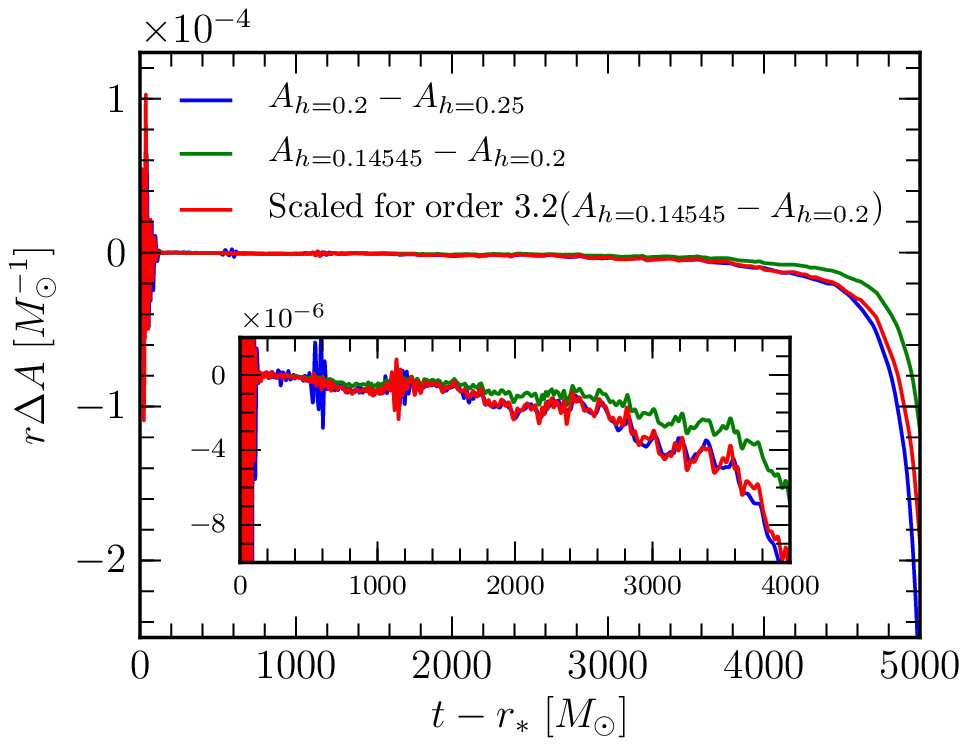}
  \end{center}
  \end{minipage}
  \begin{minipage}{\columnwidth}
  \begin{center}
    \includegraphics[width=\columnwidth]{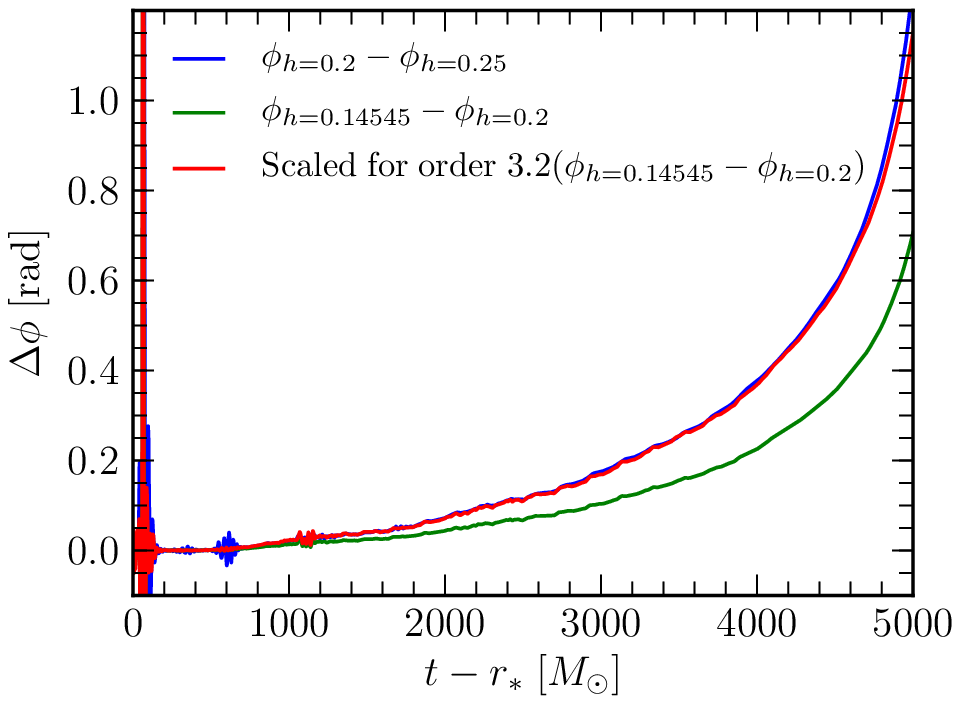}
  \end{center}
  \end{minipage}
  \caption{\label{fig:bns60.conv} Amplitude differences (left panel) and
    accumulated de-phasing (right panel) on the $\ell=2, m=2$ mode of the Weyl
    scalar $\Psi_4$ extracted at $r = 450\ M_\odot$. For both quantities
    we show the differences between the low- and the medium-resolution
    runs (blue lines), between the high- and the medium-resolution runs
    (green lines), as well as the rescaled differences between the high-
    and the medium-resolution runs (red lines) computed assuming a
    convergence order of $3.2$.}
\end{figure*}

Figure \ref{fig:bns60.ampl} shows the amplitude of the $\ell=2, m=2$ mode
of $\Psi_4$, as extracted at radius $r = 450\ M_\odot$, and as a function
of the retarded time $t - r_*$, where $r_* = r + 2 M \log ( {r}/{2 M} -
1)$.  The first thing to note is that the merger time, defined as the
time where the curvature GW amplitude $|\Psi_4|$ has its maximum, is very
close among the different runs. As we change the resolution by a factor
$1.7$ from low to high, the differences in the merger time are only of
the order of $\simeq 2.5\ \%$. In comparison, the results reported by
\citep{Hotokezaka2013b} show, for a model with the same compactness,
changes of the order of $\simeq 20\ \%$ when changing the resolution by a
factor $1.4$, despite their highest resolution being about $35\ \%$
higher than our highest one (this roughly corresponds to a factor $3$
increase in the computational costs).  Having such small differences in
the merger time allows us to perform a much simpler and cleaner analysis
with respect to the one presented in \citep{Hotokezaka2013b}. In
particular we do not need to perform any alignment/time scaling of the
numerical waveforms when measuring the convergence order or when
performing their Richardson extrapolation.

A more quantitative analysis is shown in Fig. \ref{fig:bns60.conv}, which
reports the convergence order in the amplitude $A$ (left panel) and in
the phase $\phi$ (right panel), which are defined as
\begin{equation}
  \Psi_4 \equiv A \ e^{i \phi}\,.
\end{equation}
We find very clean convergence in both quantities with order $3.2$ almost
up to the NR contact time, which we estimate following
\citep{Bernuzzi2012} to be $t - r_* = 5000\ M_\odot$ ($2 M_\infty \omega
\simeq 0.11$, where $\omega = d{\phi}/dt$). Notice that the NR contact
happens before the ``bare'' contact frequency introduced by
\citep{Damour:2012} $2 M_\infty \omega_{\mathrm{cont}} = 2
\mathcal{C}^{3/2} \simeq 0.15$, which, is instead reached at $t - r_*
\simeq 5200\ M_\odot$ in the highest resolution run and should really be
seen as an upper-limit \citep{Damour:2012}.

What is shown in Fig.~\ref{fig:bns60.conv} represents the highest
convergence order ever shown for BNS simulations in full
general-relativity. It is smaller than the nominal one of the scheme
(which is fourth since we use fourth-order finite-differencing for the
spacetime), but this is to be expected because HRSC methods typically
reach their nominal convergence order only at very high resolutions
\citep{Shu97, Radice2012a}. More importantly, and as already mentioned
above, this high order of convergence is obtained without any
manipulation of the waveforms, which is a procedure hardly justified from
a mathematical point of view, although used by some of us
\citep{Baiotti:2009gk,Read2013} and in \citep{Hotokezaka2013b}. As also
observed with other codes \citep{Bernuzzi2011}, our solution shows a loss
of convergence (super-convergence) after $t \gtrsim 5000\ M_\odot$, as
this represents the time after which the stars merge. This time is
different for different resolutions and inevitably leads to a loss of
convergence.

\section{Tidal effects}

\begin{figure}
\begin{center}
  \includegraphics[width=\columnwidth]{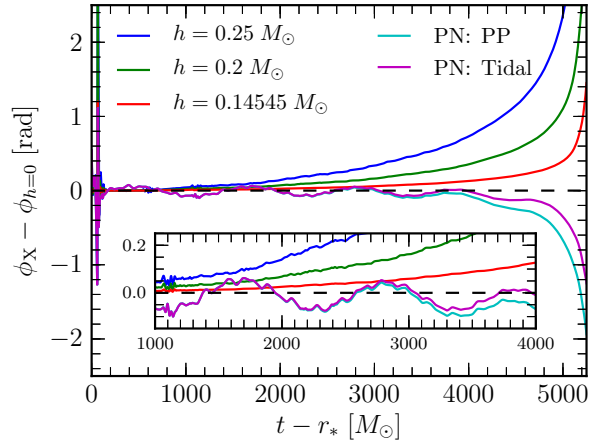}
  \caption{\label{fig:bns60.thc.vs.analytic} Accumulated de-phasing with
  respect to the Richardson-extrapolated NR waveform assuming convergence
  order of $3.2$. In particular, we show the de-phasing accumulated by
  the three simulations with increasing resolution (blue, green and red
  lines, respectively), as well as by the waveforms predicted by the PN
  Taylor-T4 approximation in the limit of PP (light-blue line) and when
  tidal corrections are included (purple line).}
\end{center}
\end{figure}

As a first direct application of our code to explore the validity of
semi-analytic approximation techniques, we perform a comparison with the
predictions from the PN theory using the Taylor-T4 formula either in the
PP approximation \citep{Santamaria2010}, or with the inclusion of tidal
effects up to relative 1PN order \citep{Hinderer09, Flanagan2008,
Vines:2010ca, Pannarale2011, Maselli2012}. This is shown in Fig
\ref{fig:bns60.thc.vs.analytic}. In particular, we take as reference the
Richardson-extrapolated phase evolution, $\phi_{h=0}$, computed using the
measured convergence order $3.2$ and we plot the de-phasing of the
different models with respect to it. Because the extrapolated waveform is
obtained using the medium and high resolutions, which do not merge up
until $t - r_* \simeq 5200\ M_\odot$, it is reasonable to extend the
comparison with the PN waveforms up to these times in Fig
\ref{fig:bns60.thc.vs.analytic}.

We align the PN waveforms in time and phase to the extrapolated one
using the $\chi^2-$minimization procedure proposed by
\citet{Boyle:2008ge}, which was also adopted by \citep{Baiotti:2010,
Baiotti2011, Bernuzzi2012, Hotokezaka2013b}. In particular we determine
$\tau$ and $\Delta\phi$ to minimize
\begin{equation}
  \chi^2 = \int_{t_1 + r_*}^{t_2 + r_*} \big[ \phi_{\mathrm{NR}}(t) -
  \phi_{\mathrm{PN}}(t - \tau) - \Delta \phi \big]^2 \dd t\,,
\end{equation}
where the interval $(t_1,t_2)$ is taken to be $(150, 2000) M_\odot$ so as
to include two local adjacent maxima of the GW phase in the early part of
the GW signal, following \citet{Hotokezaka2013b}. These local extrema are
due to the phase modulation induced by the orbital eccentricity of the
initial data and our choice of the matching interval allows us to avoid
over-fitting these modulations with the least-square procedure. The
results that we present below are not sensitive with respect to the
choice of the window for the fit as long as it is large enough to avoid
over-fitting the eccentricity phase modulation and, at the same time,
small enough so as not to include the last part of the inspiral.

When comparing among numerical solutions we find that the de-phasing
between the highest resolution run and the Richardson-extrapolated result
is of $\simeq 0.4$ radians at NR contact point, $t - r_* = 5000\ M_\odot$
(which is about $13.5$ GW cycles; see red solid line in
Fig.~\ref{fig:bns60.thc.vs.analytic}), and of $\simeq 1.4$ radians over
$\sim 15$ GW cycles at the bare contact frequency. As a comparison, for a
model with the same compactness, \citet{Hotokezaka2013b} found a
de-phasing of $\simeq 5$ radians between the highest resolution
simulation (which is even $35\ \%$ higher than ours) and the extrapolated
solution over $15$ GW cycles at the bare contact frequency. We remark,
however, that the Richardson-extrapolated waveform should be treated with
care after the NR contact, since convergence is lost then.

On the other hand, when comparing with semi-analytical predictions we
find that the de-phasing between the PP PN waveform and the extrapolated
one at $t - r_* = 5000\ M_\odot$ is only of $\simeq 0.65$ radians
(light-blue solid line). With the inclusion of tidal effects the
de-phasing is further reduced to only $\simeq 0.35$ radians, \ie to the
point that it is almost comparable to the uncertainty due to the
eccentricity of the initial data, which we  estimated to be $\pm 0.1$
radians (see inset in Fig.~\ref{fig:bns60.thc.vs.analytic}). Indeed, the
tidally corrected PN waveform appears to be very close to the
Richardson-extrapolated data up to the NR contact point, $t - r_* = 5000\
M_\odot$, with the accumulated de-phasing at the bare contact frequency
being now of only $\simeq 0.9$ radians. This result clearly rules out, at
least for this model, the importance of any significant tidal
contributions from next-to-next-to-leading--order terms in the PN
expansion. 

When comparing our results with those published recently, we note that
\citet{Bernuzzi2012} reached conclusions similar to ours, namely, that
semi-analytic approximations, such as the effective one-body (EOB) and
the tidally-corrected Taylor-T4 PN expansion, are able to describe
accurately the phasing of the binaries essentially up to contact. Their
results, however, were based on more deformable stellar models, for which
the tidal de-phasing is intrinsically larger. On the other hand, for
stellar models with smaller deformability, which is comparable but still
larger than the one considered here, \citet{Hotokezaka2013b} found that
all the available analytic models underestimate the tidal deformability
in the very last phase of the inspiral. It is thus possible that the
conclusions reached by \citet{Hotokezaka2013b} may have been influenced
by larger numerical viscosity, as the one in the early calculations of
\citet{Baiotti2011}. A direct comparison using the same stellar models
could help clarify this point.

Since we are considering Richardson-extrapolated results, our estimates
of the de-phasing need to be reported with a certain degree of error. We
can follow \citep{Hotokezaka2013b} and estimate the error assuming a
variance of $\pm 0.2$ in the convergence order used in the
extrapolation. If we do, and because of our high convergence order, we
find an error-bar that is only $\pm 0.05$ radians at $t - r_* =
5000\ M_\odot$. In practice, however, this uncertainty is below a larger
and systematic error coming from the initial eccentricity of the binary.

\section{Conclusions}

We have presented the first higher-than-second-order, multi-dimensional,
general-relativistic hydrodynamics code: \texttt{WhiskyTHC}, result of
the combination of the \texttt{Whisky} \citep{Baiotti04, Galeazzi2013}
and \texttt{THC} \citep{Radice2012a} codes.  We have applied it to the
simulation of the late-inspiral and merger of two neutron stars in
quasi-circular orbits. We showed that our code is able to accurately
estimate the small tidal effects present in the inspiral of binaries with
realistic compactness, $\mathcal{C} = 0.18$ and small tidal number
$\kappa_2 = 0.05$, at a much lower resolution and at a fraction of the
cost used so far in the literature, \eg \citep{Baiotti:2010, Baiotti2011,
Bernuzzi2012, Hotokezaka2013b}. In particular, we found a convergence
order of $3.2$ in both the amplitude and the phase of the GWs up to the
contact point in the numerical simulations.  When comparing the numerical
Richardson-extrapolated waveform with the analytic PN predictions, we
found remarkable agreement, especially when tidal corrections are
included. At least for the case considered here, our results indicate
that the tidally corrected Taylor-T4 waveform agrees very well with the
NR waveform up to contact, \ie up to frequencies, of the order of $\simeq
1\ \mathrm{kHz}$, which are significantly higher than the $450\
\mathrm{Hz}$ conservatively estimated by \citep{Hinderer09} as a validity
limit for the PN expansion considered here. For this reason we can
exclude a significant contributions from: 1) $\ell=2$ linear tidal terms
higher than 1PN order, 2) $\ell>2$ tidal terms, 3) non-linear tidal
terms.

Having developed a very accurate and high-order code, we are now ready to
exploit its efficiency to explore systematically the role of tidal
effects in BNS mergers with simple and realistic EOSs. In addition, we
will use it to test and improve semi-analytical descriptions, \eg PN and
the effective one body (EOB)~\citep{Baiotti:2010}, and to quantify the
detectability of tidal effects by advanced GW detectors, following the
spirit of~\citep{Read2013}.

\section*{acknowledgements}
It is a pleasure to acknowledge W. Kastaun for kindly providing the
primitive recovery routine, F. Pannarale for providing the PN
waveforms. We also thank D. Alic and K. Takami for numerours useful
discussions. Partial support comes from the VESF grant (EGO-DIR-69-2010),
the DFG grant SFB/Transregio 7, and by ``Compstar'', a Research
Networking Programme of the European Science Foundation. The calculations
were performed on the SuperMUC cluster at the LRZ and on the Datura
cluster at the AEI.

\bibliographystyle{mn2e.bst}
\bibliography{aeireferences}
\end{document}